\documentclass[twocolumn]{aastex62}

\usepackage{graphicx}

\newcommand{\sourcelong}{ZTFJ1539+5027}
\newcommand{\source}{ZTFJ1539}
\newcommand{\fdot}{\dot{f}_0}

\newcommand{\amp}{\mathcal{A}}
\newcommand{\Tobs}{T_{\rm LISA}}

\begin{document}

\defcitealias{ZTF}{B19}

\title{Prospects for Gravitational Wave Measurement of ZTFJ1539+5027}


\author{Tyson B. Littenberg}
\affiliation{NASA Marshall Space Flight Center, Huntsville, AL 35812, USA}
\author{Neil J. Cornish}
\affiliation{eXtreme Gravity Institute, Montana State University, Bozeman, MT 59717, USA}

\begin{abstract}
The short-period eclipsing binary \sourcelong\ discovered by \cite{ZTF} will be a strong gravitational-wave source for the Laser Interferometer Space Antenna (LISA). We study how well LISA will constrain the parameters of this system by analyzing simulated gravitational wave data, and find that LISA observations will significantly improve measurements of the distance and inclination of the source, and allow for novel constraints to be placed on the speed of gravity.
\end{abstract}


\section{\label{sec:intro}Introduction}
Ultra compact binaries (UCBs) are expected to be the most numerous gravitational wave (GW) sources in the mHz band, with tens of sources already discovered electromagnetically~\citep{Kupfer:2018}.  
The Laser Interferometer Space Antenna (LISA) is a planned space-based GW detector which will have peak sensitivity in the frequency range populated by short-period UCBs.
The expected sensitivity of LISA is such that the UCBs will dominate the data stream from 0.1 to several mHz, making them an important source for LISA science.

The discovery of \sourcelong\ \citep[][hereafter B19]{ZTF} as part of an ongoing survery by ZTF~\cite{2019PASP..131a8002B} + KPED~\cite{2019MNRAS.485.1412C} has added a new UCB system to the census of known LISA sources, adding to other recent discoveries such as SDSS J0651 + 2844~\cite{2011ApJ...737L..23B} and WD 0931+444~\cite{2014MNRAS.444L...1K}.

\sourcelong, henceforth abbreviated as \source, is among the shortest-period sources currently known. High frequency systems are ideal for precision measurement by LISA as they are in the most sensitive part of the LISA band, away from the confusion foreground due to low-frequency UCBs, and will exhibit detectable changes in the orbital period due to the loss of energy and angular momentum through GW emission, breaking parameter degeneracies and enabling detailed investigations of the orbital dynamics of the binary.

The exquisite measurement of \source\ will serve as input to multimessenger analysis of the source, combining what is now known from electromagnetic (EM) observations with what will be learned when LISA is operational. 
The information encoded in the GW signal is complementary to what can be measured electromagnetically, providing direct measures of distance and inclination, and independently constraining the orbital dynamics of the binary.  

In this letter we simulate a fiducial GW signal using the EM-measured binary parameters (randomizing over the GW observables that are unknown), and apply a Markov Chain Monte Carlo (MCMC) pipeline for the detection and characterization of UCBs in LISA data~\citep{Littenberg:2011zg,Littenberg:2019}.
We confirm that \source\ will be a loud source for LISA.
LISA observations will improve the measurement of the binary inclination with $1\sigma$ uncertainty within ${\sim}1^{\circ}$ after 1 yr of observing, and down to ${\sim}0.15^{\circ}$ by the end of an 8 yr extended mission--a factor of ${\sim}5$ improvement over the current measurement.
Assuming the orbital evolution is dominated by GW emission as predicted by General Relativity, and marginalizing over uncertainty in tidal dissipation, the distance to \source\ will be measured by LISA to a comparable level as currently known after a 2 yr observation, 
and improving the current distance measurement by a factor of 10 after an extended mission.
Improvements to the distance measure are realized within 1 yr of LISA observations if measurement of \source's orbital evolution improves due to continued EM monitoring of the source. LISA will constrain the eclipse times to a precision of $\lesssim 1$ s, allowing for a novel constraint on the speed of gravity.

\section{\label{sec:method} Method}

UCB GW signals are typically modeled with eight free parameters: GW frequency at the start of observations $f_0$, time derivative of the frequency $\fdot$; the GW amplitude $\amp$; two angles encoding the sky location ($\alpha$, $\beta$); and three angles encoding the orientation of binary's orbital angular momentum vector with respect to the observer ($\iota$, $\psi$, and $\varphi_0$) which are the inclination, polarization angle, and phase of the waveform at the start of observations. 

The simulated GW signals for this study were placed at the sky location of \source\ with parameters derived from the central values quoted in Table 1 of \citetalias{ZTF} for the orbital period and it's first time derivative $P$, $\dot{P}$; inclination $i$ (here $\iota$), and distance $d$ (here $d_L$).
The polarization and initial phase are not constrained by EM measurements, and so were drawn randomly from uniform distributions over their supported range. 
The parameter values for the simulated GW signals are as follows: 
$f_0=0.00482170$ Hz, 
$\fdot=2.76428021\times10^{-16}$ s$^{-2}$,
$\amp=1.01275314\times10^{-22}$,
$\cos\iota = 0.1$,
$\psi \in U[0,\pi]$, and
$\phi_0 \in U[0,2\pi]$.
Measurement of parameters most relevant to the understanding of \source\ are independent of the specific choice of $\psi$ and $\phi_0$, as the UCBs are continuous sources and the LISA orbital motion allows both GW polarizations to be well sampled for $\Tobs>1$ yr.  
The specific realization of $\psi$ and $\phi_0$ will effect the ``instantaneous'' signal strength over short time intervals, such as when answering the question of how quickly \source\ will become detectable.



The waveform model is efficiently calculated in the frequency domain using a fast-slow decomposition described in~\cite{Cornish:2007if}.
We simulate LISA data to be consistent with the instrument sensitivity described in~\citet{LISA_PROPOSAL}, using the orthogonal time delay interferometry (TDI) channels $A$ and $E$~\citep{Tinto2014}.

UCBs exhibit little frequency evolution and are thus contained within a narrow bandwidth, a feature exploited in data analysis strategies for processing the full galaxy~\citep{Crowder_2006eu, Littenberg:2011zg}.
Our analysis will focus only on the narrow frequency band of data containing \source\ assuming there are no other sources present. 
At GW frequencies $\gtrsim 4$ mHz it is expected that sources are sufficiently separated in frequency that source confusion will not be an issue, although other sources will likely be present in the analysis bandwidth used here.
To illustrate we show the LISA response to a full galaxy simulation consistent with those used for the LISA Data Challenges~\citep{LDC} adapted from \cite{Korol:2017}, and have added \source\ for context.
Data are simulated for different LISA observing times to quantify how the GW measurement evolves over the mission.
$\Tobs=1$, 2, 4, and 8 yr are presented, with 4 yr being a proxy for the nominal mission lifetime, and 8 yr as a possible extended mission. 

Source parameters are measured using the open source trans-dimensional MCMC code under development as a prototype algorithm for LISA signal processing~\citep{ldasoft}.
The signal model is fixed to the sky location of \source\, i.e. we are conducting a targeted search for the source.

Taking full advantage of joint observations requires the multivariate posterior distribution from the electromagnetic observations be mapped to the GW parameters for use as priors.
As a proxy for when and how joint multi-messenger analyses will improve what is understood about \source\, we look to when the GW observations become informative given what is currently known about the source. 
To do so we adopt uniform priors over the remaining free GW parameters to quantify the circumstances under which LISA observations are competitive with, or surpass, the measurement precision obtained in \citetalias{ZTF}.

Posterior samples from the MCMC are post-processed into marginalized distributions for detailed investigations of particular parameters.
To compare properties constrained by the EM measurements using the GW observables, we re-parameterize samples from the MCMC to derive posterior distributions conditional on the underlying assumptions for which the mappings are valid.

Marginalized 1D posteriors are shown using Kernel Density Estimation (KDE) to smooth the finite-sampling of the distributions.
The GW posteriors are compared to the measurements of the source parameters by the EM observations, which are represented as Gaussian distributions with central values and standard deviations taken from Table 1 of \citetalias{ZTF}.
For the inclination, where the errors are asymmetric, we take the average of the upper and lower interval as the standard deviation. 


\section{Results}\label{sec:results} 

\source\ is a strong LISA source, as reported in \citetalias{ZTF}, with signal-to-noise ratio of ${\sim}140$ at $\Tobs\sim4$ yr.
By simulating the source with random polarization angle and initial phase (unconstrained by EM observations) we find $\mathcal{O}(20\%)$ of simulations led to detection in one week of data when targeting the specific sky location of \source.
This is largely a curiosity, as the source will unambiguously be detectable after $\Tobs\sim\mathcal{O}(1)$ month regardless of the initial conditions.

According to population synthesis simulations of the galaxy (e.g.~\cite{Toonen:2012, Korol:2017}) the source is in a richly populated part of the LISA measurement band, but above the frequencies where the superposition of UCBs blend together to form a confusion-limited astrophysical foreground~\citep{Bender:1997}.
We find for the nominal LISA mission lifetime that 99\% of sources in the 4-6 mHz band have maximum overlaps with any other source of less than 50\%, and a median maximum overlap of less than 0.03, indicating that \source\ has a low probability of being mis-identified or blended with other sources. Consequently, the presence of other sources is unlikely to impact our results.

\begin{figure}[ht]
\centering
\includegraphics[width=\linewidth]{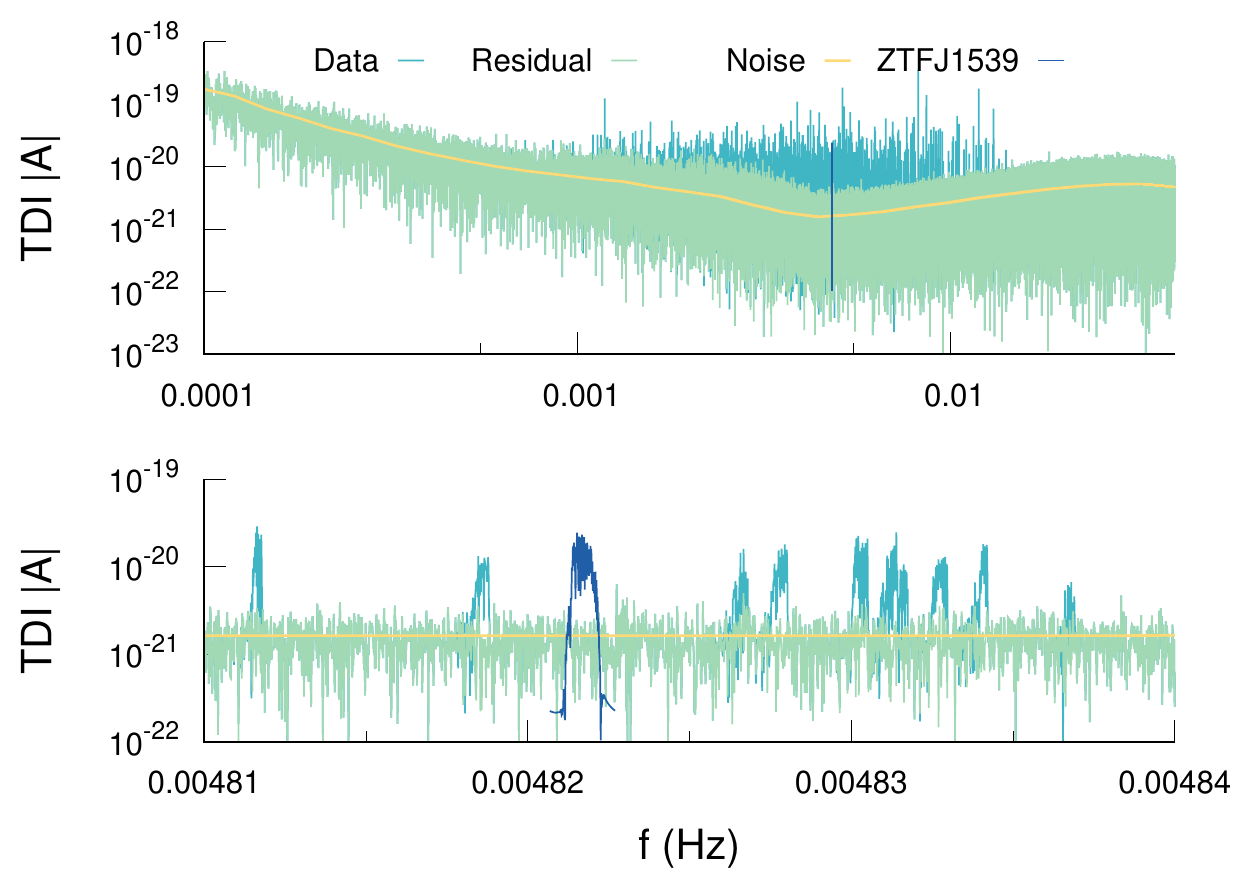}
\caption{\small{Example LISA data containing sources from a simulated galaxy (blue), and the residual after identifiable sources have been removed (green), with \source\ added for context (dark blue). 
The source frequency is above the confusion noise (top panel) and has low probability of being confused with other UCBs (bottom panel).}}
\label{fig:spectra}
\end{figure}

Fig.~\ref{fig:spectra} shows the power spectral density of the TDI $A$ data stream of the simulated LISA response to a Milky Way-like population of UCBs (blue) after $\Tobs\sim2$ yr.
After the resolvable systems are subtracted from the data, the residual spectrum (green) shows a bump between ${\sim}0.4$ and ${\sim}4$ mHz due to the confusion noise.
The yellow line is a fit to the RMS noise level, including instrument and confusion noise.  
As shown in the upper panel, \source\ (dark blue) is among the brighter sources in the LISA band, at higher frequency than the confusion noise.
The lower panel shows a narrower frequency range which demonstrates that, while densely populated, individual sources like \source\ are clearly identifiable.


\begin{figure}[ht]
   \centering
   \includegraphics[width=\linewidth]{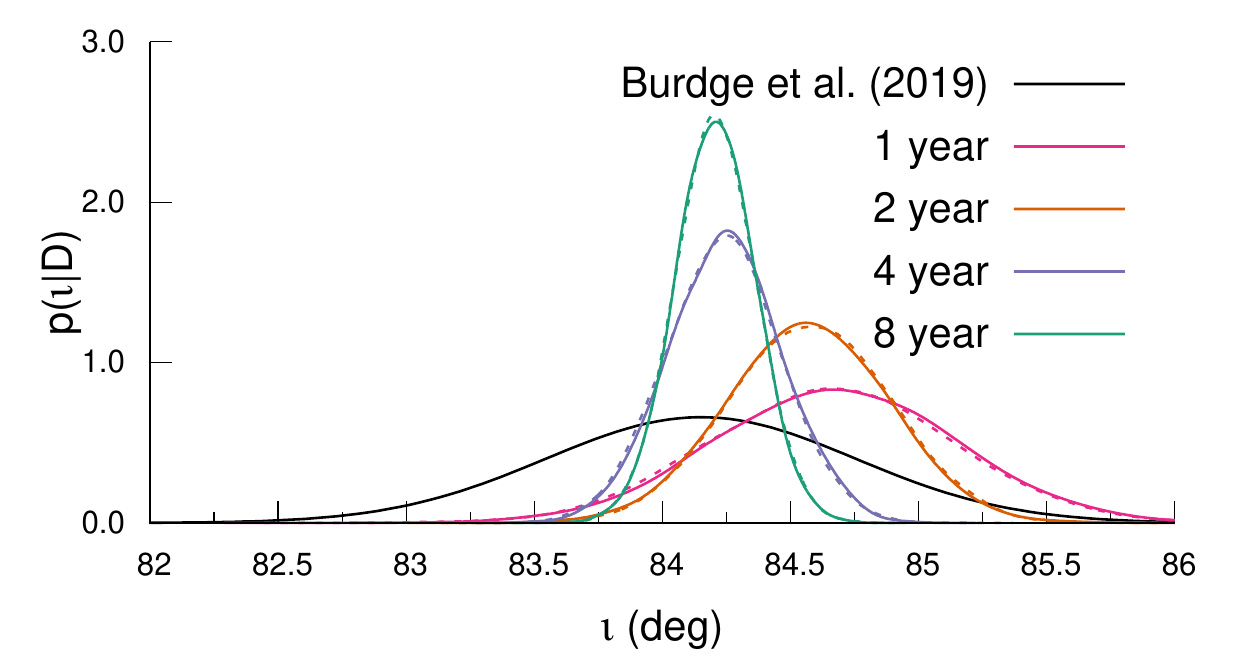} 
   \includegraphics[width=\linewidth]{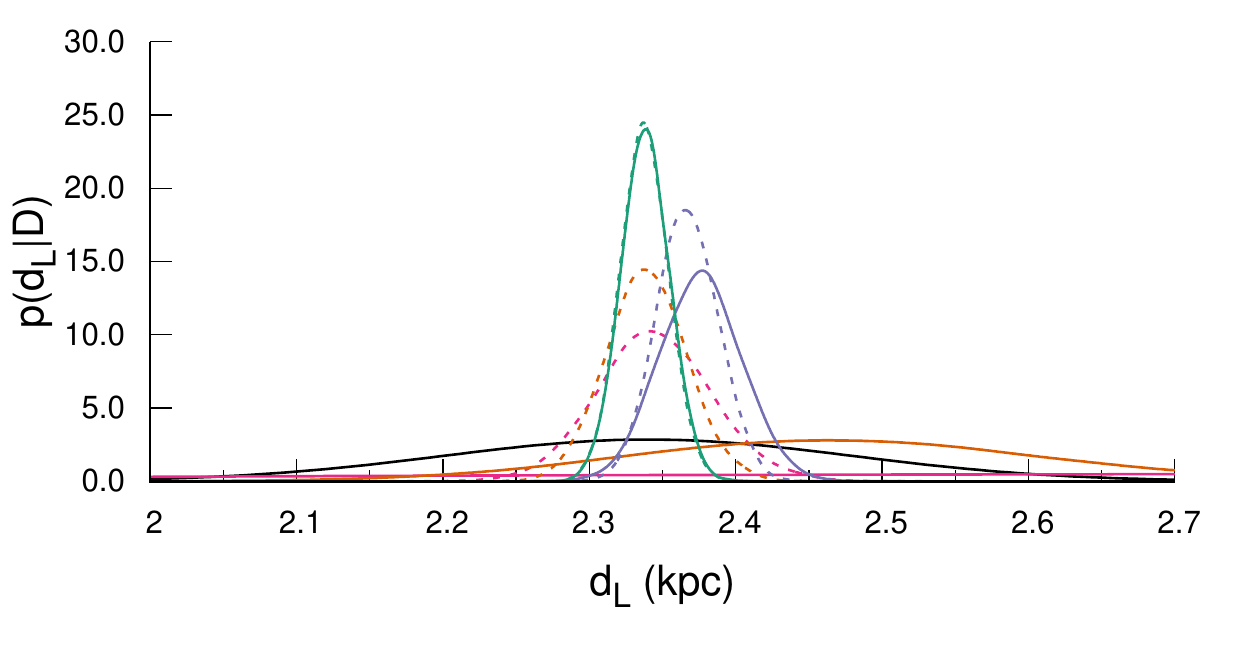} 
   \caption{\small{GW-derived constraints on the inclination (top) and distance to the binary (bottom) as a function of $\Tobs$, from 1 yr (magenta) to 8 yr (teal). Solid curves use GW-only information, dashed curves use priors on $f_0$ and $\fdot$ derived from the EM observations.}}
   \label{fig:inc_and_distance}
\end{figure}

Comparing the timing of the eclipses between EM and GW observations allows us to place novel constraints on the speed of gravity~\cite{Cutler:2002ef}. Measurements of the GW initial phase $\phi_0$ map to uncertainties in the timing of the eclipses: $\delta t = \delta\phi_0/(2\pi f_0)$. Dividing by the light travel time from the system we can constrain the speed of gravity to be within $3.2 \; (2.1)\times 10^{-12}$ of the speed of light. While a factor of $\sim1000$ weaker than from the joint GW-EM observations of GW170817~\citep{Monitor:2017mdv}, the constraints are at very different frequencies and in very different environments. This is particularly relevant for alternative theories of gravity that invoke a screening mechanism~\cite{Creminelli:2017sry}.

The inclination is encoded in the GW signal by the relative amplitudes of the two polarization modes $h_+$ and $h_\times$. 
Due to the orbital motion of the LISA constellation, the detector samples both polarizations through the year enabling a direct constraint on inclination.
The upper panel of Fig.~\ref{fig:inc_and_distance} shows the marginalized posterior distributions on inclination given the GW data $D$, $p(\iota|D)$, as a function of $\Tobs$.
The LISA constraint on inclination will already improve on what is currently known after $\Tobs=1$ yr.
The inclination constraint continues to improve throughout the mission, reaching $0.2^\circ$ ($0.15^\circ$) precision by the end of the nominal (extended) mission.

For UCBs that exhibit GW-driven orbital evolution, the distance to the source is derived from the frequency evolution and GW amplitude. Tidal interactions in the binary are an additional source of dissipation which needs to be accounted for when mapping the GW parameters to distance.

\begin{figure}[ht]
   \centering
   \includegraphics[width=\linewidth]{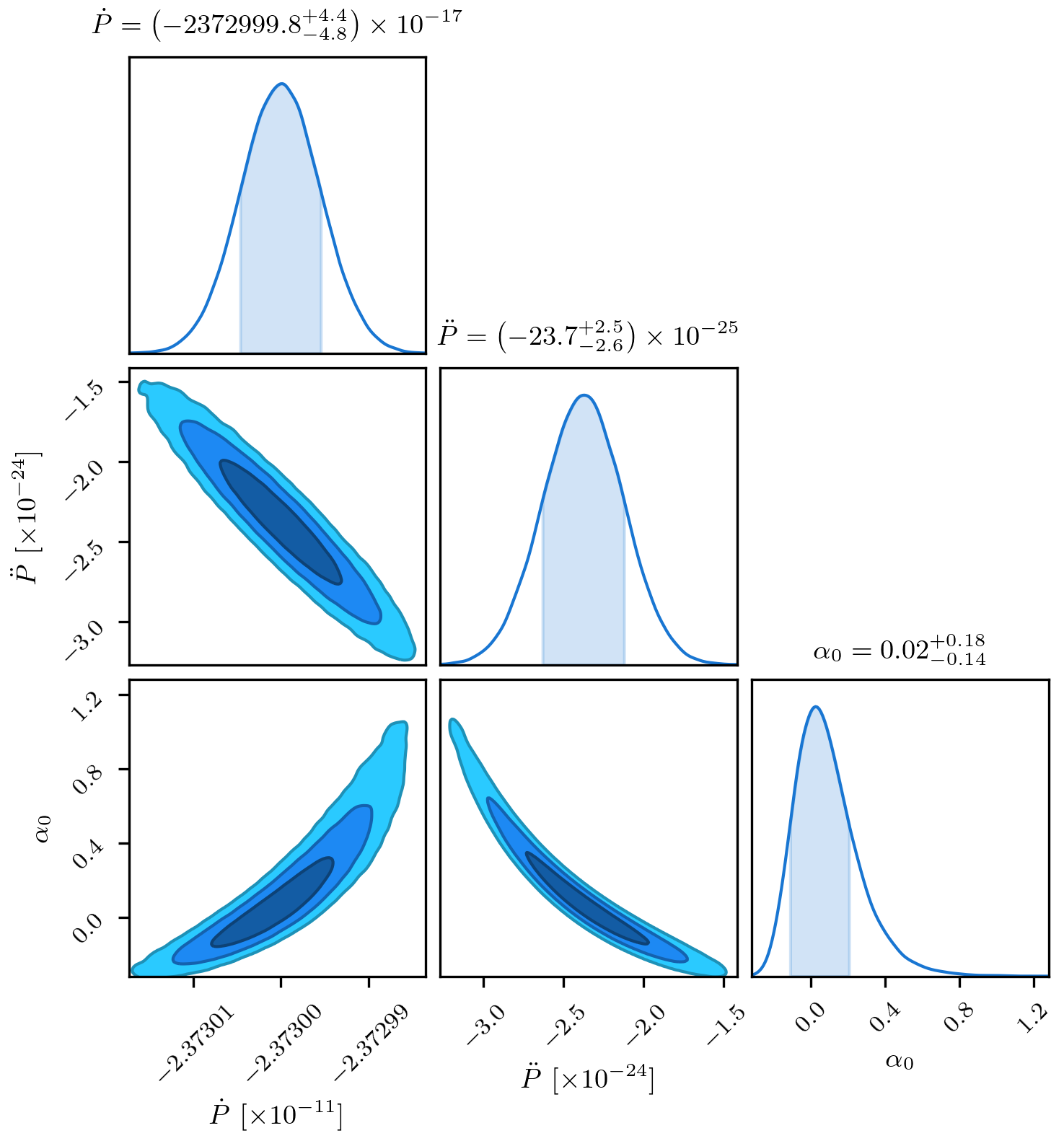} 
   \caption{\small{Projected constraints on the period evolution and tidal enhancement parameter $\alpha_0$ with an additional 20 yrs of EM observations.}}
   \label{fig:P-Pdot}
\end{figure}

To leading post-Newtonian order, the amplitude is $\amp = 2 {\cal M}^{5/3}(\pi f_0)^{2/3}/d_L$, where ${\cal M}$ is the chirp mass.
Measurements of the individual masses from \citetalias{ZTF} constrain ${\cal M}$ to within 6\%, but a more precise measurement can be derived from the rate of orbital decay. Allowing for tidal contributions, the frequency evolves as $\dot f = (1+\alpha_0\, (f/f_0)^{4/3})\dot f_{\rm GW}$, where $\dot f_{\rm GW}$ is the usual GW driven decay and $\alpha_0$ is the tidal enhancement factor. Using these relations to eliminate ${\cal M}$, we find $\amp = 5 \fdot/(48\pi^2(1+\alpha_0) f_0^3 d_L)$. The analysis of \citetalias{ZTF} estimates the tidal enhancement to be $\alpha_0 = 0.067\pm 0.005$. This estimate is subject to unquantified theoretical uncertainties, so it is preferable to measure $\alpha_0$  directly. In principle this can be done by measuring the second derivative of the period, $\ddot P$. To see if this is feasible, we took the existing eclipse data from \citetalias{ZTF} and simulated an additional 20 years of once yearly observations by a large telescope
capable of timing the eclipses to $\sim10$ ms. Note that the measurement errors on the period evolution scale linearly with the timing accuracy, and as the square root of the timing cadence: once yearly measurements with large instruments capable of 10 ms timing are equivalent to nightly measurements at 0.2 s accuracy using smaller instruments such as KPED.
The results of the simulation are shown in Fig.~\ref{fig:P-Pdot} using the ZTF derived period evolution parameters and tidal enhancement factor. We found that, while $\ddot P$ can be constrained at the 4\% level, $\alpha_0$ is poorly measured. Thus, the most precise limits on $\alpha_0$ will continue to come from the types of modeling described in \citetalias{ZTF}.

We remap the MCMC samples on $(f_0,\fdot,\amp)$ to produce $p(d_L|D)$, marginalizing over $\alpha_0$ using Eq. 9 and Table 1 of \citetalias{ZTF} and drawing from 1D Gaussian distributions for the binary parameters.  We find no difference in the results when assuming $10\%$ or $1\%$ errors on the $\kappa_i$ constants which encode the internal structure of the white dwarfs. Measurement of the radii and, to a slightly lesser extent, the masses contribute most to the uncertainty in $\alpha_0$.

The lower panel of Fig.~\ref{fig:inc_and_distance} compares $p(d_L|D)$ with results in~\citetalias{ZTF}. 
Displayed are inferred distributions when only the GW data is considered (solid curves), or using priors on $f$ and $\fdot$ assuming continued improvement in the measurement of $P$ and $\dot{P}$ as shown in Fig.~\ref{fig:P-Pdot} (dashed).
LISA observations of \source\ will improve the distance measurement by $\Tobs\sim 1$ yr when incorporating projected constraints from the EM observations, and within 4 yr using the GW data only.  
The advantage gained by incorporating the EM priors diminishes as $\Tobs$ increases because the measurement becomes dominated by the uncertainty in $\amp$.
Ultimately, the distance measurement precision improves by a factor of 10 for $\Tobs\gtrsim4$ yr, localizing the source to within $\mathcal{O}(10)$ pc at 1$\sigma$.

\section{\label{sec:discuss} Discussion}
The discovery of \source\ adds to the census of known UCBs detectable through GW emission.
The source will be unambiguously identified by LISA owing both to its inherently large GW amplitude and
its location in the GW spectrum where source confusion is not likely to complicate the analysis.

GW measurements will independently provide comparable levels of precision to the current measurement of the orbital evolution of the system, and will improve the precision to which the source location and orientation are known.
Measurement of the inclination of the binary's orbital plane is immediately improved within the first year of observations, achieving a factor of 5 decrease in uncertainty. 
The GW constraint on the distance to the source is a more remarkable improvement, reducing uncertainty by a factor of 10, localizing \source\ to within 10 pc. Multi-messenger observations of \source\ will allow the speed of gravity in the mHz regime to be constrained to parts in a trillion.


Not addressed in this analysis is how the LISA measurements will feed back into the EM observations to further improve understanding of the source.
Multi-messenger observations are greater than the sum of their parts, and a true joint analysis incorporating both the EM and GW data, exploiting the complex relationships between observables and theoretical models, will further reveal what will be achieved when studying UCBs in the LISA era.

With over a decade until the LISA launch, and the advent of large-scale ground-based survey instruments like \emph{Gaia}, ZTF, and LSST, the number of known GW sources will continue to grow providing increasingly more opportunities for conducting joint analyses of UCBs.

\acknowledgments
We thank Q. Baghi, J. Slutsky, and J.I. Thorpe for drawing our attention to the discovery paper. We thank K. Burdge and M. Coughlin for answering questions about future EM observing campaigns for \source.
TBL acknowledges the support of NASA grant number NNH15ZDA001N and the NASA LISA Study Office. NJC acknowledges the support of NASA grant number 80NSSC19K0320.

\bibliographystyle{aasjournal}
\bibliography{sources}

\end{document}